\documentclass[aps,amsmath,amssymb,amsfonts,showpacs,twocolumn,pra]{revtex4}
\usepackage{amscd}
\usepackage{amsthm}
\usepackage{amsmath}
\usepackage{graphicx}
\usepackage{array}
\usepackage{cases}

\usepackage[
  colorlinks,
  linkcolor = blue,
  citecolor = blue,
  urlcolor = blue]{hyperref}

\def\qed{\hfill \vrule height7pt width 7pt depth 0pt}

\def\ra{\rangle}

\def\be{\begin{equation}}
\def\ee{\end{equation}}
\def\ba{\begin{array}}
\def\ea{\end{array}}

\begin{document}
\title{Connecting the UMEB in $\mathbb{C}^{d}\bigotimes\mathbb{C}^{d}$ with partial Hadamard matrices}
\author{Yan-Ling Wang$^{1}$, Mao-Sheng Li$^{2}$, Shao-Ming Fei$^{3, 4}$, Zhu-Jun Zheng$^{1}$}

 \affiliation
 {
   {$^1$Department of Mathematics,
 South China University of Technology, Guangzhou
510640, P.R.China} \\
{$^2$Department of Mathematical Sciences,
 Tsinghua University, Beijing
100084, China}\\
{$^3$School of Mathematical Sciences, Capital Normal University,
Beijing 100048, China}\\
{$^4$Max-Planck-Institute for Mathematics in the Sciences, 04103
Leipzig, Germany}
}

\begin{abstract}

We study the unextendible maximally entangled bases (UMEB) in $\mathbb{C}^{d}\bigotimes\mathbb{C}^{d}$
and connect it with the partial Hadamard matrix. Firstly, we show that for a given special UMEB in $\mathbb{C}^{d}\bigotimes\mathbb{C}^{d}$,
there is a partial Hadamard matrix can not extend to a complete Hadamard matrix in $\mathbb{C}^{d}$.
As a corollary,  any $(d-1)\times d$ partial Hadamard matrix can extend to a complete Hadamard matrix. Then we obtain that for any $d$ there is an UMEB except $d=p\ \text{or}\ 2p$, where $p\equiv 3\mod 4$ and $p$ is a prime. Finally,  we argue that there exist different kinds of constructions of UMEB in $\mathbb{C}^{nd}\bigotimes\mathbb{C}^{nd}$ for any $n\in \mathbb{N}$ and $d=3\times5 \times7$.
\end{abstract}

\pacs{03.67.Hk,03.65.Ud }\maketitle
\maketitle

\section{Introduction}
It is well known that the quantum states are divided into two parts: separable states and entanglement states. The pure product states are special case of the separable states while the maximally entangled states play important roles for  the entangled states\cite{nils,EofCon,Be,mhor}. One of the significant property of the quantum theory is the quantum nonlocality. An  unextendible product bases (UPBs) in bipartite quantum system {\small$\mathbb{C}^{m}\bigotimes\mathbb{C}^{n}$} is a set of orthogonal product states less than $mn$ such that no further product states are orthogonal to every state in that set\cite{BD2,BD1}. It is proven that the UPBs display some nonlocality without entanglement\cite{BD2,De}. Similar with the UPBs,
in 2009, S. Bravyi and J. A. Smolin first proposed the notion of  unextendible maximally entangled basis(UMEB): a set of orthonormal
maximally entangled states in {\small$\mathbb{C}^{d}\bigotimes\mathbb{C}^{d}$} consisting of fewer than $d^2$ vectors which have no
additional maximally entangled vectors that are orthogonal to all of them.  The authors pointed out that the UMEBs are helpful  for  constructing some quantum states with special property of the entanglement of assistance(EOF) and can be used to find
quantum channels that are unital but not convex mixtures of unitary operations \cite {s3}.

It is proved that there do not exist UMEBs
for $d=2$, and  a 6-member UMEB for $d=3$ and a 12-member UMEB for $d=4$  were constructed \cite {s3}.
After that, the construction of UMEB has attracted the attention some authors. Firstly,  there are many UMEBs have been constructed in $\mathbb{C}^{d}\bigotimes\mathbb{C}^{d'} (d\neq d')$\cite{BC,Li}. In Ref.\cite{Wang}, the authors studied the  UMEB in $\mathbb{C}^{d}\bigotimes\mathbb{C}^{d}$, and gave that if there is an UMEB in $\mathbb{C}^{d}\bigotimes\mathbb{C}^{d}$ then there is also an UMEB in $\mathbb{C}^{qd}\bigotimes\mathbb{C}^{qd}$ for any $q\in \mathbb{N}$.
Nevertheless, for  the UMEB in $\mathbb{C}^{d}\bigotimes\mathbb{C}^{d}$ we only know the cases $d=3,4,3n, 4n$. So it is interesting to consider the UMEB in other higher-dimensional system $\mathbb{C}^{d}\bigotimes\mathbb{C}^{d}$ for general $d$. In addition, Guo generalized the UMEB problem by replacing the condition of maximality of states with states of given Schmidt number\cite{Guo,Guo2}.

The construction of Hadamard matrix is also an interesting topic.  In this paper, we mainly concern about the complex Hadamard matrix. The authors who are interested in complex Hadamard matrix may look  Refs.\cite{Hadam1,Hadam2,Hadam3} for further reading.  A partial Hadamard matrix is a matrix $H\in M_{m\times n}(\mathbb{T})$(where $\mathbb{T}=\{z\in \mathbb{C}\ \  \big|\ \ |z|=1\} $), whose rows are pairwise orthogonal. Given a partial Hadamard matrix $H\in M_{m\times n}(\mathbb{T})$ one interesting problem is that of deciding whether this matrix extends or not to an $n\times n$ complex Hadamard matrix. In the real case, there are many results\cite{Hadam4,Hadam5}. But for the general complex case, however, very little  seems to be known about this question\cite{Hadam6}.

In this paper, we show a relation between these two basic concepts, and in particular we show that if there are a special UMEB in $\mathbb{C}^{d}\bigotimes\mathbb{C}^{d}$, then we can find a corresponding partial Hadamard matrix which can not be extended to a complete  Hadamard matrix, and vice versa. Then by using the extendibility of any $d^2-1$ orthogonal maximally entangled states, we give an  answer to  the conjecture in \cite{Hadam6}.
The above relation between  UMEB and  partial Hadamard matrix  gives us a method construct UMEB. As a example,
we first construct a 23-member UMEB in {$\mathbb{C}^{5}\bigotimes\mathbb{C}^{5}$.
Then we generalized the example to higher dimensions:
we show that for any $n\in \mathbb{N}$, there exists an UMEB in {$\mathbb{C}^{4n+1}\bigotimes\mathbb{C}^{4n+1}$}.
At last,  we show that there exists an UMEB in  {$\mathbb{C}^{d}\bigotimes\mathbb{C}^{d}$} except $d=p\ \text{or}\ 2p$, where $p\equiv 3\mod 4$ and $p$ is a prime. In addition, we also give an UMEB in $\mathbb{C}^{7}\bigotimes\mathbb{C}^{7}$ for the exceptional  unsolved cases. Then by using  the UMEBs constructed from $d=3,5,7$, we show there are different kinds of UMEBs in $\mathbb{C}^{(3\times 5\times 7)n}\bigotimes\mathbb{C}^{(3\times 5\times 7)n}$ for any $n\in \mathbb{N}.$
\section{The UMEBs in $\mathbb{C}^{d}\bigotimes\mathbb{C}^{d}$ and the partial Hadamard matrix}

\noindent{\bf Definition 1.} A set of states \{$|\phi_{a}\rangle\in\mathbb{C}^{d}\bigotimes\mathbb{C}^{d}:\,a=1,2,\cdots,n,\,n<d^2$\}
is called an $n$-number UMEB if and only if
(i) $|\phi_{a}\rangle$, $a=1,2,\cdots,n$, are maximally entangled;
(ii) $\langle\phi_{a}|\phi_{b}\rangle=\delta_{ab}$;
(iii) if $\langle\phi_{a}|\psi\rangle=0$ for all $a=1,2,\cdots,n$, then $|\psi\rangle$ cannot be maximally entangled.

Here under computational basis a maximally entangled state $|\phi_{a}\rangle$ can be expressed as
\be\label{1}
|\phi_{a}\rangle=(I\otimes U_a)\,\frac{1}{\sqrt{d}}\sum_{i=1}^d|i\ra\otimes |i\ra,
\ee
where $I$ is the $d\times d$ identity matrix, $U_a$ is any unitary matrix. According to (\ref{1}),
a set of unitary matrices {$\{U_a\in M_d(\mathbb{C})|a=1,...,n\}$} gives an $n$-number UMEB in {$\mathbb{C}^{d}\bigotimes\mathbb{C}^{d}$} if and only if\\
 (i)  $n<d^2$;\\
 (ii) {$Tr(U_a^\dag U_b)=d\,\delta_{ab},~~ \forall  a,b=1,\cdots,n$;\\}
 (iii) For any {$U\in M_d(\mathbb{C}),$} if {$ Tr(U_a^\dag U)=0,~\forall\,  a=1,\cdots,n$,} then {$U$} cannot be unitary.

 In this paper, we only use the latter equivalent  difinition of UMEB.

\noindent{\bf Definition 2.}\cite{Hadam6} Partial Hadamard matrices: A partial Hadamard matrix in $\mathbb{C}^n$ is a rectangular matrix $H$ with entries in the circle $\mathbb{T}$ whose rows are pairwise orthogonal. That is,  $H\in M_{m\times n}(\mathbb{T})(m<n)$, and $HH^\dagger=nI_m$.

\noindent{\bf Definition 3.}\cite{Hadam6} We call a partial Hadamard matrice  $H\in M_{m\times n}(\mathbb{T})$ in $\mathbb{C}^n$ is completable if there exists a Hadamard matrix $\widetilde{H}$ whose first $m$ rows equal to the rows of $H$ respectively.

\medskip
\noindent{ \textit{Lemma} 1.} If there is an $N$-number UMEB in {\small$\mathbb{C}^{d}\bigotimes\mathbb{C}^{d}$}, then for any $q\in\mathbb{N}$,
there is a $\widetilde{N}$-number, $\widetilde{N}=(qd)^2-(d^2-N)$, UMEB in {$\mathbb{C}^{qd}\bigotimes\mathbb{C}^{qd}$\cite{Wang}.}

\medskip

In this paper, we mainly study with the UMEB in $\mathbb{C}^{d}\bigotimes\mathbb{C}^{d}$ contaning the following set and we call it   a special   UMEB if exists.
$$S_0=\{X^mZ^n\mid m=1,2\ldots,d-1, n=0,1,\ldots,d-1\}$$ where $X=\displaystyle\sum_{j=0}^{d-1}|j+1\rangle\langle j|,\  Z=\displaystyle\sum_{j=0}^{d-1}\omega_d^j|j\rangle\langle j|, \ \omega_d=e^{\frac{2\pi i}{d}}.$

Suppose $A=(a_{st})_{k\times d}$ is a $k\times d$ partial Hadamard matrix in $\mathbb{C}^d$, $\alpha_1,\alpha_2,\ldots,\alpha_k$ are rows of $A$.
then  we can  construct a set of unitary matrices , denoted by $S(A)=\{diag(\alpha_s)\mid s=1,2,\ldots,k\},$ where $diag(\alpha_s)=\displaystyle\sum_{t=1}^{d}a_{st}|t-1\rangle\langle t-1|.$

Then the elements in $S(A)$ are unitary and orthogonal with each other under inner product $\langle A,B\rangle=Tr(AB^{\dagger}).$

\medskip
\noindent{ \textit{Proposition} 1.} Given a partial Hadamard  matrix $A$ with $k\times d$, then $S_0\cup S(A)$ can not be extended to a full maximally  entangled base (MEB) if and only if $A$ can not extend to a complete Hadamard matrix.

\noindent{\emph{Proof.}} $\Rightarrow:$ Suppose $A$ can extend to a complete Hadamard matrix. That is there are $d-k$ mutually orthogonal vectors $\nu_1,\nu_2,\ldots,\nu_{d-k}$ with modules $1$ for each entry which are orthogonal to all rows of $A$. Then $\{U_j=diag(\nu_j) \ \big| j=1,2,...,d-k\}$ are unitary matrices which are orthogonal with each other, and lies in the orthogonal complement of $S_0\cup S(A)$. Then $S_0\cup S(A)\cup \{U_j\mid j=1,2,\ldots,d-k\}$ is a MEB. This is contradicted with $S_0\cup S(A)$ can not be extended to MEB.

$\Leftarrow:$ If $S_0\cup S(A)$ can extend to MEB, then there are $d-k$ orthogonal matrices $U_1,U_2,\ldots,U_{d-k}$ which lie in $(S_0\cup S(A))^{\bot}$.
However, $S_0^{\bot}$ is the set of diagonal matrices. Hence, $(S_0\cup S(A))^{\bot}\subseteq S_0^{\bot}$ is a subset of diagonal matrices. Suppose
$U_j=diag(\nu_j)$ for some vector $\nu_j$ in $\mathbb{C}^d$ for each $j\in \{1,2,\ldots,d-k\}$. Then the unitary of the matrix $U_j$ gives that the entries of $\nu_j$ are all module $1$. The orthogonality of $S(A)\cup \{U_1,U_2,\ldots,U_{d-k}\}$ give that {\footnotesize $\left(
                                                                            \begin{array}{c}
                                                                              A \\
                                                                              \nu_1 \\
                                                                              \nu_2 \\
                                                                              \vdots \\
                                                                              \nu_{d-k} \\
                                                                            \end{array}
                                                                          \right)$}
is a Hadamard matrix.\qed

\medskip
Now we give an  answer to  the conjecture in \cite{BD1} which conjecture that  any partial Hadamard matrix of $4\times 5$ can be complemented to a complete Hadamard matrix.

\medskip
\noindent{ \textit{Corollary 1}.} If $d$ is an integer, $d\geq 2$, and $A$ is a partial Hadamard matrix of $(d-1)\times d$. Then $A$ can be complemented to a complete Hadamard matrix.

\noindent{ \emph{Proof 1.}} Since $A$ is a $(d-1)\times d$ matrix, then we have $S_0\cup S(A)$ is a set of maximally entangled states with $d^2-1$ states. By \cite{s3}, it can be extended to MEB. Hence by Proposition 1, $A$ can be complemented to a Hadamard matrix.

\noindent{ \emph{Proof 2.}} Suppose {\small $A=\left(
                                     \begin{array}{c}
                                       \alpha_1 \\
                                       \alpha_2\\
                                       \vdots \\
                                       \alpha_{d-1} \\
                                     \end{array}
                                   \right)
$},
then we have
\begin{equation*}
\begin{array}{l}
  \text{dim}_{\mathbb{C}}(\text{span}_{\mathbb{C}}\{\alpha_1,\alpha_2,\ldots,\alpha_{d-1}\}) = d-1, \\[1.5mm]
  \text{dim}_{\mathbb{C}}(\text{span}_{\mathbb{C}}\{\alpha_1,\alpha_2,\ldots,\alpha_{d-1}\})^{\bot} = 1.
\end{array}
\end{equation*}
Choosing a nonzero vector $$\nu_d=(\nu_{d1},\nu_{d2},\ldots,\nu_{dd})\in\\ \text{span}_{\mathbb{C}}\{\alpha_1,\alpha_2,\ldots,\alpha_{d-1}\}^{\bot}$$ such that $\|\nu_d\|=1$. Then {\small $U=\left(
                                                                                                        \begin{array}{c}
                                                                                                          \frac{\alpha_1}{\sqrt{d}} \\
                                                                                                          \frac{\alpha_2}{\sqrt{d}}  \\
                                                                                                          \vdots \\
                                                                                                          \frac{\alpha_{d-1}}{\sqrt{d}}  \\
                                                                                                          \nu \\
                                                                                                        \end{array}
                                                                                                      \right)
$}
is a matrix with normal rows and orthogonal with each other.That is, $U$ is an unitary matrix. Then all columns of $U$ are also normal and orthogonal with each other.  Hence, $|\nu_{dk}|=\frac{1}{\sqrt{d}}$ for $k=1,2,\ldots,d$. Then {\small $\left(
                                                                                                        \begin{array}{c}
                                                                                                          \alpha_1\\
                                                                                                          \alpha_2 \\
                                                                                                          \vdots \\
                                                                                                          \alpha_{d-1}\\
                                                                                                          \sqrt{d}\nu_d \\
                                                                                                        \end{array}
                                                                                                      \right)
$}
is a Hadamard matrix.\qed

\medskip
\noindent{\bf Remark 1:} The Proposition 1  give us a method to construct some sets of UMEB. Suppose there is a partial Hadamard matrix $A$ whose orthogonal complement contains no vector with each entry module $1$. Then $S_0\cup S(A)$ is an UMEB.

\medskip

\noindent{ {\textit{Example} 1.}} In $\mathbb{C}^{5}\bigotimes\mathbb{C}^{5}$, there exists an UMEB with $23$ elements.

Let { $A=\left(
                 \begin{array}{c}
                   \alpha_1 \\
                   \alpha_2 \\
                 \end{array}
               \right)
=\left(
\begin{array}{ccccc}
           1 & 1 & 1 & 1 & 1 \\
           1 & -1 & 1 & \omega & \omega^2 \\
\end{array}
\right)$,}
where $\omega=e^{\frac{2\pi i}{3}}.$ If we denote \begin{eqnarray*}
                                                    \nu_1 &=& (\frac{1}{\sqrt{2}},0,-\frac{1}{\sqrt{2}},0,0), \\
                                                    \nu_2 &=& (\frac{1}{\sqrt{10}},0,\frac{1}{\sqrt{10}},\frac{2\omega^2}{\sqrt{10}},\frac{2\omega}{\sqrt{10}}), \\
                                                    \nu_3 &=& (0,\sqrt{\frac{3}{5}},0,\frac{\omega-1}{\sqrt{15}},\frac{\omega^2-1}{\sqrt{15}}),
                                                  \end{eqnarray*}
then $\text{span}_{\mathbb{C}}\{\alpha_1,\alpha_2\}^{\bot}=\{\nu_1, \nu_2, \nu_3\}.$
Let $\alpha=k_1\nu_1+k_2\nu_2+k_3\nu_3$ is a vector with each entries module $1$, that is
\begin{equation*}
\begin{cases}
|\frac{k_1}{\sqrt{2}}+\frac{k_2}{\sqrt{10}}|=1,\\[1.2mm]
|-\frac{k_1}{\sqrt{2}}+\frac{k_2}{\sqrt{10}}|=1,\\[1.2mm]
|\sqrt{\frac{3}{5}}k_3|=1,\\[1.2mm]
|k_1|^2+|k_2|^2+|k_3|^2=1.
\end{cases}
\end{equation*}
\bigskip

Then from the above equations,  we have $|k_1|=|k_2|=|k_3|=\sqrt{\frac{5}{3}}$. Moreover, $k_2=\pm ik_1, k_3=\pm ik_1.$

If we let $\alpha_3=\sqrt{\frac{5}{3}}\nu_1+i\sqrt{\frac{5}{3}}\nu_2+i\sqrt{\frac{5}{3}}\nu_3$, then {\small B=$\left(
                                                                                                      \begin{array}{c}
                                                                                                        A \\
                                                                                                        \alpha_3 \\
                                                                                                      \end{array}
                                                                                                    \right)
$} is also a partial Hadamard matrix. However, any vector lies in $\text{span}_{\mathbb{C}}\{\alpha_1,\alpha_2,\alpha_3\}^{\bot}\subseteq \text{span}_{\mathbb{C}}\{\alpha_1,\alpha_2\}^{\bot}$. Hence, if $\nu\in \text{span}_{\mathbb{C}}\{\alpha_1,\alpha_2,\alpha_3\}^{\bot}$ with each entry module $1$, then $\nu$ can be written as the form $$\nu=k_1\nu_1\pm ik_1\nu_2\pm ik_1\nu_3$$
However, $\nu$ can not be orthogonal with $\alpha_3$. Hence, the set $S_0\cup S(B)$  is an UMEB with 23 elements in $\mathbb{C}^{5}\otimes\mathbb{C}^{5}.$ \qed

\bigskip
\noindent{\textit{ Proposition} 2.} In $\mathbb{C}^{4n+1}$, there exists a partial Hadamard matrix which can not complete to a  Hadamard matrix.

\noindent \emph{Proof:}  Let

\bigskip
\begin{widetext}

{ $$A=\left(
              \begin{array}{c}
                \alpha_1 \\
                \alpha_2 \\
                \alpha_3 \\
                \vdots \\
                \alpha_{2n} \\
              \end{array}
            \right)
=\left(
                               \begin{array}{cccccccccc}
                                 1 & 1 & 1 &\cdots & 1 & 1 &1 &1 & \cdots& 1 \\
                                 1 & \omega &\omega^2& \cdots & \omega^{2n-1} & 1 &\sigma& \sigma^2 & \cdots & \sigma^{2n} \\
                                 1 & \omega^2 & \omega^4 &\cdots& \omega^{2(2n-1)} & 1 & \sigma^2 &\sigma^4& \cdots & \sigma^{2(2n)} \\
                                 \vdots & \vdots & \vdots & \vdots & \vdots & \vdots & \vdots & \vdots & \vdots & \vdots \\
                                 1 & \omega^{2n-1} & \omega^{2(2n-1)} & \cdots & \omega^{(2n-1)(2n-1)} & 1 & \sigma^{2n-1} &\sigma^{2(2n-1)}& \cdots & \sigma^{2n(2n-1))} \\
                               \end{array}
                             \right)
$$}
\end{widetext}
where $\omega=e^{\frac{2\pi i}{2n}},\  \sigma=e^{\frac{2\pi i}{2n+1}}.$
Firstly, we compute the orthogonal complement of the subspace $V$ spanned by the rows of $A.$
Obviously, {\begin{eqnarray*}
 \beta_1&=& (\frac{1}{\sqrt{2}},\overbrace{0,0,\cdots,0}^{2n-1},-\frac{1}{\sqrt{2}}, \overbrace{0,0,\cdots,0}^{2n} ), \\
 \beta_2 &=&(\frac{1}{\sqrt{8n+2}},\overbrace{0,0,\cdots,0}^{2n-1},\frac{1}{\sqrt{8n+2}},\frac{2\sigma^{2n}}{\sqrt{8n+2}}, \\
 & &\frac{2\sigma^{2n-1}}{\sqrt{8n+2}},\cdots,\frac{2\sigma}{\sqrt{8n+2}} )
           \end{eqnarray*}}
are orthogonal with $\alpha_1, \alpha_2, \ldots, \alpha_{2n}$ and $\beta_1\bot\beta_2.$ Now we set
\begin{eqnarray*}
  \gamma_1 &=& (\overbrace{0,1,0,0,\cdots,0,0}^{2n},\overbrace{0,0,\cdots,0}^{2n+1}), \\
 \gamma_2 &=& (0,0,1,0,\cdots,0,0,0,0,\cdots,0), \\
    &\vdots & \\
  \gamma_{2n-1} &=& (0,0,0,0,\cdots,1,0,0,0,\cdots,0).
\end{eqnarray*}
By Schmidt  orthogonalization we have
{\small
\begin{eqnarray*}
  \beta_3 &=& (\overbrace{0,\sqrt{\frac{2n+1}{4n+1}},0,0,\ldots,0,0,0}^{2n+1},\beta_{31},\beta_{32},\cdots,\beta_{3,2n}), \\
  \beta_4 &=& (0,0,\sqrt{\frac{2n+1}{4n+1}},0,\ldots,0,0,0,\beta_{41},\beta_{42},\cdots,\beta_{4,2n}), \\
  \  &\vdots& \\
  \beta_{2n+1} &=& (0,0,0,0,\ldots,0,\sqrt{\frac{2n+1}{4n+1}},0,\beta_{2n+1,1},\cdots,\beta_{2n+1,2n}).
\end{eqnarray*}
}
Then we have $\text{span}_{\mathbb{C}}\{\alpha_1,\alpha_2,\ldots,\alpha_{2n}\}^{\bot}=\text{span}_{\mathbb{C}}\{\beta_1,\beta_2,\ldots,\beta_{2n+1}\}.$
Suppose $\nu=k_1\beta_1+k_2\beta_2+\cdots+k_{2n+1}\beta_{2n+1}$ is a vector with entries module $1$. Then we must have
\begin{equation*}
\begin{cases}
                |\frac{1}{\sqrt{2}}k_1+\frac{1}{\sqrt{8n+2}}k_2| = 1, \\[1.2mm]
                |\frac{1}{\sqrt{2}}k_1-\frac{1}{\sqrt{8n+2}}k_2| = 1, \\[1.2mm]
                |\sqrt{\frac{2n+1}{4n+1}}k_3| = 1, \\[1.2mm]
                |\sqrt{\frac{2n+1}{4n+1}}k_4| = 1, \\[1.2mm]
                \ \ \ \ \ \ \ \vdots \\[1.2mm]
                |\sqrt{\frac{2n+1}{4n+1}}k_{2n+1}| = 1, \\[1.2mm]
                |k_1|^2+|k_2|^2+\cdots+|k_{2n+1}|^2=|\nu|^2 = 4n+1.
              \end{cases}
\end{equation*}

Solving the above equations, we have $$|k_1|=|k_2|=\cdots=|k_{2n+1}|=\sqrt{\frac{4n+1}{2n+1}},\text{and}\  k_1=\pm ik_2.$$
If $A$ can be extended to a  Hadamard matrix, by adding $2n+1$ rows $\nu_1,\nu_2,\ldots,\nu_{2n+1}.$ Then we have
\begin{eqnarray*}
  \nu_1 &=& k_{11}\beta_1+k_{12}\beta_2+\cdots+k_{1,2n+1}\beta_{2n+1}, \\
  \nu_2 &=& k_{21}\beta_1+k_{22}\beta_2+\cdots+k_{2,2n+1}\beta_{2n+1}, \\
  &\vdots  & \ \ \ \ \ \ \ \ \ \ \  \ \\
   \nu_{2n+1} &=& k_{2n+1,1}\beta_1+k_{2n+1,2}\beta_2+\cdots+k_{2n+1,2n+1}\beta_{2n+1}.
\end{eqnarray*}
The above analysis gives that $|k_{st}|=\sqrt{\frac{4n+1}{2n+1}}$. Clearly, the orthogonality of $\nu_1,\nu_2,\ldots,\nu_{2n+1}$ give that vectors
$(k_{11},k_{12},\ldots,k_{1,2n+1}),(k_{21},k_{22},\ldots,k_{2,2n+1}),\ldots,\\(k_{2n+1,1},k_{2n+1,2},\ldots,k_{2n+1,2n+1})$ are orthogonal with each other.
Hence, if we let $K=(k_{st})_{(2n+1)\times(2n+1)}$. Then $\sqrt{\frac{2n+1}{4n+1}}K$ is a matrix with entries module $1$ and each row are mutually orthogonal. Hence, $H=\sqrt{\frac{2n+1}{4n+1}}K$ is a Hadamard matrix.
In the following we show that this can not be true and get a contradiction.

 If we replace each row by $(k_{j1},k_{j2},\ldots,k_{j,2n+1})$ by $\frac{1}{k_{j1}}(k_{j1},k_{j2},\ldots,k_{j,2n+1}),$
then the new matrix $\widetilde{H}$ is also a Hadamard matrix with the element of first column all are $1$. Noticing that $k_{j2}=\pm ik_{j1}$ for $j=1,2,\cdots,2n+1$ so   the elements in  second column of $\widetilde{H}$ are $i$ or $-i$.
The Hadamard matrix $\widetilde{H}$ also give that the column of $\widetilde{H}$ are orthogonal with each other. Suppose there are $p$ elements of the sencond column are $i$ and $q$ elements are $-i$.  Then the inner product of the first column and the second column is $(p-q)i$. Here $p+q=2n+1$, so $p\neq q$. Hence, $(1,1,\ldots,1)^{T}$
can not orthogonal with the second column.

Hence we can conclude that $A$ can not be extended to a  Hadamard matrix.\qed

\medskip
\noindent{ \textit{Corollary} 2.} There exists an UMEB in $\mathbb{C}^{4n+1}\otimes \mathbb{C}^{4n+1}$ for any integer $n$.

\medskip
\noindent{ \textit{Corollary} 3.} There exists an UMEB in $\mathbb{C}^{d}\otimes \mathbb{C}^{d}$, whenever $d\neq p\ \text{or}\ 2p\ $($p\equiv 3\ \text{mod} \ 4$ and $p$ is a prime).

\noindent{\emph{Proof:}} Let $d=p_1^{r_1}p_2^{r_2}\ldots p_k^{r_k}\ \text{where}\ p_i $ are primes,  $p_1<p_2<...<p_k,$ and $r_i\in \mathbb{N}$ for $i=1,2,\ldots,k.$
If  $p_1=2, r_1\geq 2$ then we have an UMEB for $d$ is multiple of 4. Else if some $p_j=4n+1(n\in \mathbb{N})$, from the corollary 2 we have an UMEB. So we can suppose that all the primes  are of the for $p_j=4n+3$ except if the first one to be 2. Now suppose there are two primes $p_j=4n+3$ and $p_s=4m+3(m\in \mathbb{N})$, then we can get $4t+1|p_jp_s$ for some integer $t$, we also can get an UMEB. Then we can get only the situation $d=p$ or $2p$, where $p=3\ \text{mod}\ 4 $ and $p$ is a prime are not solved. \qed

\medskip

 We have solved the most situations, only the cases $d=p\ \text{or}\ 2p$, where $p=3\ \text{mod}\ 4 $ and $p$ is a prime are not solved. Among all the numbers $d$ which are unsolved, 7 is the smallest one. In the following, we sovle this case by the same method.

\medskip
\noindent{ \textit{Example} 2.} In $\mathbb{C}^{7}\otimes\mathbb{C}^{7}$, there exists an UMEB with $45$ elements.

Let { $A=
\left(
\begin{array}{c}
 \alpha_1 \\
  \alpha_2 \\
   \alpha_3 \\
    \end{array}
    \right)
=\left(
\begin{array}{ccccccc}
  1 & 1 & 1 & 1 & 1 & 1 & 1 \\
   1 & \omega & \omega^2 & 1 & i & -1 & -i \\
    1 & \omega^2 & \omega & 1 & -1 & 1 & -1 \\
       \end{array}
        \right)$}

where $\omega=e^{\frac{2\pi i}{3}}$, and Obviously,
{ \begin{eqnarray*}
\beta_1&=&(\frac{1}{\sqrt{2}},0,0,-\frac{1}{\sqrt{2}},0,0,0),\\ \beta_2&=&(\frac{1}{\sqrt{14}},0,0,\frac{1}{\sqrt{14}},\frac{-2i}{\sqrt{14}},\frac{-2}{\sqrt{14}},\frac{2i}{\sqrt{14}}), \\ 
\end{eqnarray*}}
are orthogonal with $\alpha_1, \alpha_2, \alpha_{3}$ and $\beta_1\bot\beta_2,$ Now we set
$$\gamma_1=(0,0,0,0,0,1,0). $$
By Schmidt  orthogonalization we have
$$\beta_3=(0,\frac{2\omega}{\sqrt{14}},\frac{2\omega^2}{\sqrt{14}},0,\frac{-i}{\sqrt{14}},\frac{2}{\sqrt{14}},\frac{i}{\sqrt{14}}). $$
$\beta_3$ is orthogonal with all the vectors $\alpha_1,\alpha_2,\alpha_3,\beta_1,\beta_2$.
Let $\beta_4$ be a normalized vector and orthogonal with $\alpha_1,\alpha_2,\alpha_3,\beta_1,\beta_2,\beta_3$.
Then we can get
$$\beta_4=(0,\frac{2}{\sqrt{14}},\frac{-2\omega^2}{\sqrt{14}},0,\frac{\omega-1}{\sqrt{14}},0,\frac{\omega-1}{\sqrt{14}}).$$
Hence we obtain that $\text{span}_{\mathbb{C}}\{\alpha_1,\alpha_2,\alpha_3\}^{\bot}=\text{span}_{\mathbb{C}}\{\beta_1,\beta_2,\beta_3,\beta_4\}$.
Suppose $\alpha=k_1\beta_1+k_2\beta_2+k_3\beta_3+k_4\beta_4$ is a vector with each entries module $1$, that is
\begin{eqnarray*}
\begin{cases}
|\frac{k_1}{\sqrt{2}}+\frac{k_2}{\sqrt{14}}|=1,\\[1mm]
|-\frac{k_1}{\sqrt{2}}+\frac{k_2}{\sqrt{14}}|=1,\\[1mm]
|{\frac{2\omega k_3}{\sqrt{14}}}+\frac{2k_4}{\sqrt{14}}|=1,\\[1mm]
|{\frac{2\omega^2 k_3}{\sqrt{14}}}-\frac{2\omega k_4}{\sqrt{14}}|=1,\\[1mm]
|\frac{-2ik_2}{\sqrt{14}}-\frac{ik_3}{\sqrt{14}}+\frac{(\omega-1)k_4}{\sqrt{14}}|=1,\\[1mm]
|\frac{2ik_2}{\sqrt{14}}+\frac{ik_3}{\sqrt{14}}+\frac{(\omega-1)k_4}{\sqrt{14}}|=1,\\[1mm]
|\frac{-2k_2}{\sqrt{14}}+\frac{2k_3}{\sqrt{14}}|=1,\\[1mm]
|k_1|^2+|k_2|^2+|k_3|^2+|k_4|^2=7.
\end{cases}
\end{eqnarray*}
Then from the first two equations above, we obtain the first two equations below. And from the third and fourth equations above, we obtain the third and fourth  equations below. So are to  the fifth and sixth. Here for two complex number $z_1=x_1+iy_1, z_2=x_2+iy_2$, we write  $z_1\perp z_2$ by meaning that $x_1x_2+y_1y_2=0$.
\begin{eqnarray*}
\begin{cases}
\frac{|k_1|^2}{2}+\frac{|k_2|^2}{14}=1,\\[1mm]
k_1\perp k_2,\\[1mm]
\frac{4|k_3|^2}{14}+\frac{4|k_4|^2}{14}=1,\\[1mm]
\omega k_3\perp k_4\Rightarrow k_3\perp \omega^2 k_4,\\[1mm]
\frac{|(\omega-1)k_4|^2}{14}+\frac{|2ik_2+ik_3|^2}{14}=1,\\[1mm]
(\omega-1)k_4\perp(2ik_2+ik_3)\Rightarrow \omega^2k_4\perp (2k_2+k_3),\\[1mm]
|\frac{-2k_2}{\sqrt{14}}+\frac{2k_3}{\sqrt{14}}|=1,\\[1mm]
|k_1|^2+|k_2|^2+|k_3|^2+|k_4|^2=7.
\end{cases}
\end{eqnarray*}
Then from the above equations, we have $|k_2|^2={\frac{7}{4}},
\frac{|2k_2+k_3|^2}{14}+\frac{3}{14}|k_4|^2=1, |k_3|^2+|k_4|^2=\frac{7}{2}, |k_2-k_3|^2=\frac{7}{2} $. Since $\omega^2 k_4\perp k_3, \omega^2k_4\perp (2k_2+k_3)$, we can get $\omega^2 k_4\perp 2k_2$, then we have $k_2, k_3$ are $\mathbb{R}$ linear dependence. So we can suppose $k_2=rk_3$, for some real number $r$. Substituting this into  above four equations, we get the following two equantions:
 \begin{eqnarray*}
\begin{cases}
r^2-r+\frac{1}{2}=0,\\[1mm]
r^2-2r-1=0.\\[1mm]
\end{cases}
\end{eqnarray*}
Then $r$ is unsolvable. So there is no $k_1,k_2,k_3,k_4$ satisfing the condition.
That is, there is no vector in $\text{span}_{\mathbb{C}}\{\alpha_1,\alpha_2,\alpha_3\}^{\bot}$ with each entry module $1$.
 Hence, by Proposition 1, we have an UMEB in $\mathbb{C}^{7}\otimes\mathbb{C}^{7}.$\qed

Actually, when $d=3$ the UMEB contains $6$ states, so it misses $3$ states to form a full base. Similarly, when $d=5$ there are $2$ states missing, when $d=7$ there are $4$ states missing. Then there are three ways to obtain  the UMEBs for $d=3\times 5\times 7$ by the method of lemma, respectively from $d=3, 5,7.$  The one obtained from $d=3$ is missing  $3\times 35=105$  states. The one obtained from $d=5$ is missing  $2\times 21=42$ states, while the last one obtained from $d=7$ is missing $4\times 15=60$. So the three UMEBs are different with each other. Hence for the case $d=3\times 5\times 7$,  there are at least three UMEBs. Moreover, it can be generalized to the case $d=3\times 5\times 7\times n$ for any integer $n$.
\section{conclusion and Discussion}
We study the UMEBs in $\mathbb{C}^{d}\bigotimes\mathbb{C}^{d}$ and connect it with the partial Hadamard matrix. We show that the existence of a special UMEB in $\mathbb{C}^{d}\bigotimes\mathbb{C}^{d}$  is equivalent to the existence of an uncompletable partial Hadamard matrix. In particular, as a corollary, we get any $(d-1)\times d$ partial Hadamard matrices can always extend to a complete Hadamard matrix, which  gives an answer to the conjecture in \cite{Hadam6}. Actually, the Proposition 1 also give us a method to construct UMEB by using an uncompletable partial Hadamard matrix. Then we prove that there exists an uncompletable partial Hadamard matrix for $d=4n+1$ which implies the  existence of  an UMEB in $\mathbb{C}^{4n+1}\bigotimes\mathbb{C}^{4n+1}$. At last, combining the lemma with the proposition 2,  we obtain that for any $d$ there is an UMEB except $d=p\ \text{or}\ 2p$, where $p\equiv 3\mod 4$ and $p$ is a prime.
In addition, we also give an UMEB by the partial Hadamard method when $d=7$. We conclude  there are at least three different sets of UMEBs in $\mathbb{C}^{d}\bigotimes\mathbb{C}^{d}$ when  $d$ is multiple of $3\times 5\times 7$.

We hope that the paper will be helpful both for the construction of UMEB and the partial Hadamard matrices.

\vspace{2.5ex}
\noindent{\bf Acknowledgments}\, \,
This work is supported by the NSFC 11475178, NSFC 11571119 and NSFC 11275131.

\end{document}